\documentclass{cimento}
\usepackage{tabularx}

\usepackage{graphicx}  

\title{Low contrast detection and super-resolution in CT images: evaluation of a novel approach based on Centroidal Voronoi Tessellation.}
\author{Lorenzo Lasagni}
\instlist{\inst{Department of Physics and Astronomy, University of Florence, Florence, Italy,} lorenzo.lasagni@unifi.it}

\begin{document}

\maketitle

\begin{abstract}
In this work, image analysis techniques used in astrophysics to detect low-contrast signals have been adapted in the processing of Computed Tomography (CT) images, combining Centroidal Voronoi Tessellation (CVT) and machine learning techniques. Several CT acquisitions were performed using a phantom containing cylindrical inserts of different diameters producing objects with different contrasts respect to background. The images of the phantom, tilted by a known angle with respect to the tomograph axis (to mimic the casual orientation of a clinical lesion), were acquired at various radiation doses ($ CTDI_{vol} $) and at different slice's thicknesses. The success in detecting the signal in the single image (slice) was always greater than 60\%. The axis of each insert has always been correctly identified. A super-resolution 2D image was then generated by projecting the individual slices of the scan along this axis, thus increasing the CNR of the object scanned as a whole. CVT holds great promise for future use in medical imaging, for the identification of low-contrast lesions in homogeneous organs, such as the liver.
\end{abstract}

\section{Introduction}
The detection of low-contrast lesions in computed tomography (CT) images is a critical task in the diagnosis and management of many diseases, including cancer, but it remains a major challenge due to limitations of imaging technology and variability of lesion appearance ~\cite{intro:uno}. A variety of image processing techniques have been developed to improve the detection of such lesions, but they often suffer from high false positive rates and poor sensitivity, particularly for small or subtle lesions~\cite{intro:uno, intro:due, intro:tre}.\\
In this article, we present a novel approach for the detection of low-contrast lesions in CT images using centroidal Voronoi tessellation (CVT). CVT is a computational geometry technique that is employed in digital images analysis to cluster regions of similar intensity. Numerous domains and applications, including astrophysical image processing, data analysis, chemical processes, cellular biology, and statistics, have adopted the CVT model. ~\cite{intro:quattro, intro:cinque, intro:sei, intro:sette, intro:otto, intro:nove}. However, according to our knowledge, a CVT-based approach had never been applied to CT images to perform lesion detection.\\ 
We exploit CVT for the detection of low contrast lesions of small or mild sizes. In addition, we developed an algorithm that, following the clustering CVT process, exploits the 3D information of the CT images to create a super-resolution image with enhanced contrast lesions. \\
The CVT-based algorithm was then validated on a series of CT images acquired on a specifically designed phantom containing cylindrical inserts, intended to simulate low-contrast lesions in human tissue. The phantom was scanned at different Volumetric Computed Tomography Dose Index ($ CTDI_{vol} $) to replicate various levels of noise and image quality ~\cite{intro:dieci}. Furthermore the phantom images were arranged at a tilted angle with respect to the CT acquisition axis, to simulate a spatial orientation of the lesion in a hypothetical patient's abdomen. The capability of the inserts detection algorithm and of their axes resulted remarkably high that we decided to ask the authorization from the ethics committee for the application of our procedure on actual clinical images.\\ 

\section{Materials and Methods}
\subsection{Phantom and CT images}
\begin{figure}[h!]
\centering
\includegraphics[scale=0.5]{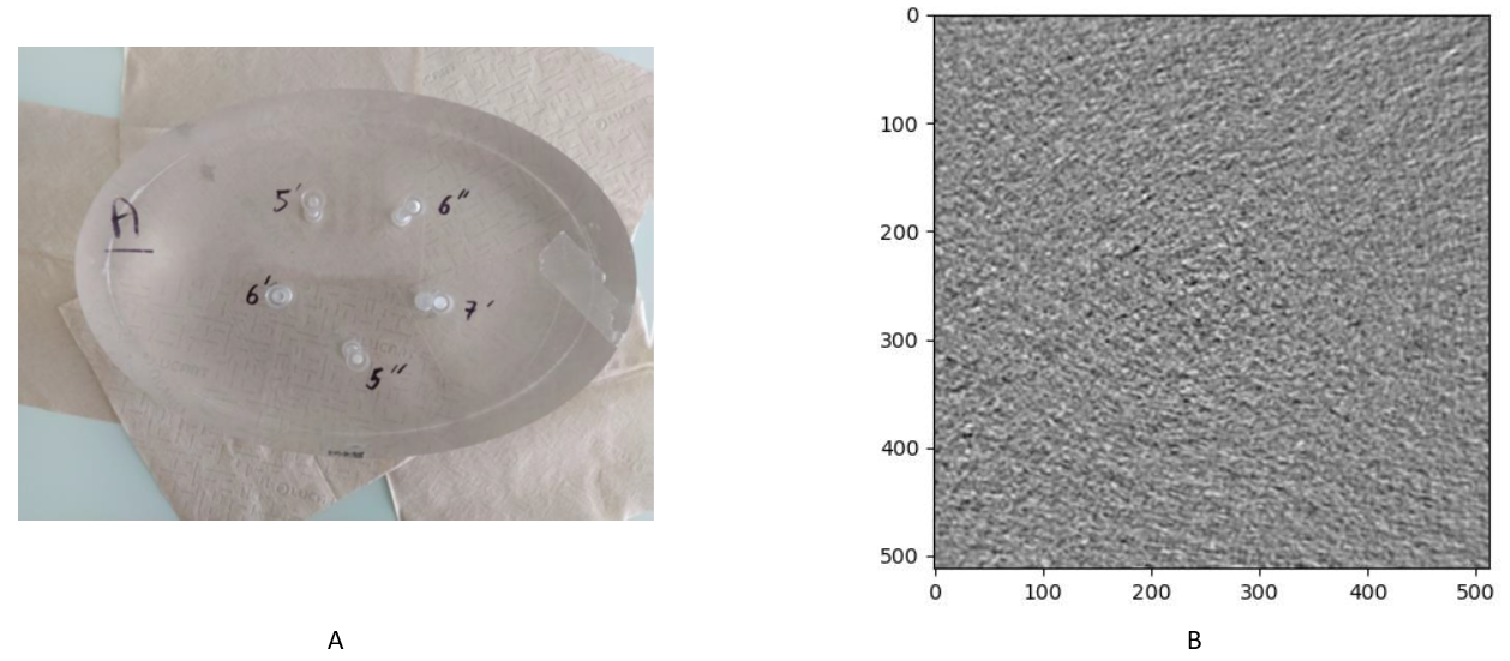}
\caption{A: picture of the phantom; B: a phantom CT slice on which the algorithm runs.}\label{fig:fantoccio}
\end{figure}

The image dataset used to test and optimize the proposed algorithm has been obtained by a specifically designed PolyMethyl MethAcrylate phantom~\cite{matmed:undici}, containing 5 cylindrical inserts of different diameters (5 mm x2, 6 mm x2, 7 mm); the inserts were filled with iodine contrast media at different concentration in order to obtain different contrast values (8, 25 and 35 Hounsfield Unit (HU) \footnote{HU is the unit used in CT to measure the density of the various tissues: 1 HU is  1/1000 the density of water}) with respect to PMMA background (125 HU at 120 kVp). Acquisitions were performed at two different $CTDI_{vol}$ settings (3.54 and 7.05 mGy), having selected the standard oncological protocol for abdomen (120 kVp, AEC, helical mode, pitch = 1, collimator aperture = 38.4 mm, slice thickness = 0.6 abd 1 mm). CT images reconstruction FoV (RFoV) was chosen equal to $15 cm^{2}$ (matrix of 512×512 pixels) in order to generate images with the highest spatial resolution containing all the five inserts. The phantom was positioned with its axis perpendicular to XY plane rotated about 16 degrees with respect to the CT image acquisition Z axis detail.
The figure \ref{fig:fantoccio} shows a picture of the phantom and of a reconstructed CT slice cointainingcontaining almost invisible low-contrast objects.
 
\subsection{The Algorithm}
\begin{figure}[h!]
\centering
\includegraphics[scale=0.65]{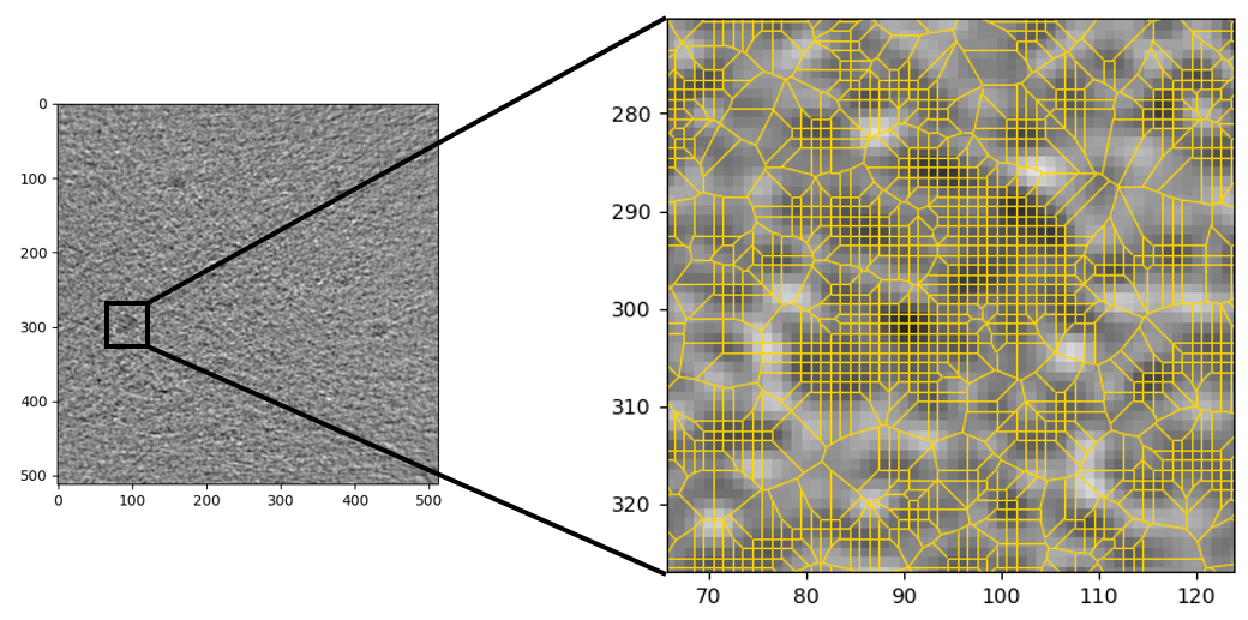}
\caption{Voronoi tessellation after the application of the MacQueen's algorithm. It is noticeable that the tessels areas are smaller in the high intenisty regions and larger outside.}\label{fig:tassvor}
\end{figure}

The first step involves the generation of 26000 points or seeds, corresponding to about 1/10 of the total number of pixels; the coordinates (x,y) of each seed are distributed within the image in a quasi-random way. This is achieved by using a Monte Carlo simulation where the most favored coordinates are picked within image areas characterized by elevated pixel intensity values (a possible signal area). The software used for the Monte Carlo simulation and the corresponding density probability definition were developed in-house. The following step consists in running 500000 total iterations of the MacQueen CVT algorithm ~\cite{intro:sei, matmed:dodici}, which iteratively updates the seeds position, only one per iteration, by moving each seed halfway towards its closest seed. Therefore the position of each seed is updated about 20 times. The accumulation of the seeds in regions of higher signal corresponds to the minimum of a cost function that locally measures the sum of squared distances between each seed and its nearest one. Since the constraint for the CVT algorithm is that each seed must be located in the mass centroid of its Voronoi tessel~\cite{intro:quattro}, it follows that seeds, the accumulation generates smaller Voronoi tessels in the regions with higher signal intensity. A representation of the Voronoi tessellation obtained after MacQueen's algorithm is shown in figure \ref{fig:tassvor}.\\

The following step involves the use of an unsupervised clustering algorithm, DBSCAN (Density-Based Spatial Clustering of Applications with Noise) ~\cite{matmed:tredici}, to identify clusters. DBSCAN is a density-based clustering algorithm for detecting clusters of arbitrary shape. It works by grouping seeds that are close together while ignoring points that are far away each other or isolated. It requires two parameters, the searching radius (eps) and the minimum number of points (MinPts), to control the density of the clusters: if a point has at least MinPts (we opted to use the lowest seed count depicted in the insert as the lower threshold, assuming that they were randomly generated) neighbors within a distance of eps, then it is considered as a core point and a new cluster is formed. In our algorith eps was chosen equal to the radius of the smallest insert, while as MinPts we have chosen the average number of pixels that are present in the low-constrast object area. DBSCAN is performed after the removal of the Voronoi seeds associated with Voronoi tessels areas larger than 10 pixels: this is the average size of the initially randomly generated tessels. A first binary mask is generated, by the output of DBSCAN, to visually identify the clusters. A second, indipendent, binary mask is generated  by agglomerating the neighboring pixels with signal values above average value of the image after a Gaussian smoothing plus $2\sigma$. The application of this smoothing technique enables improved homogeneity and spatial coherence of signal regions, which may otherwise be influenced by incidental noise fluctuations. A logical AND is then performed between the two masks. This intersection allows to remove the possible noise's clusters erroneously selected from the two previous single steps,thus making the identification of signal clusters more robust. \\

\begin{figure}[h!]
\centering
\includegraphics[scale=0.5]{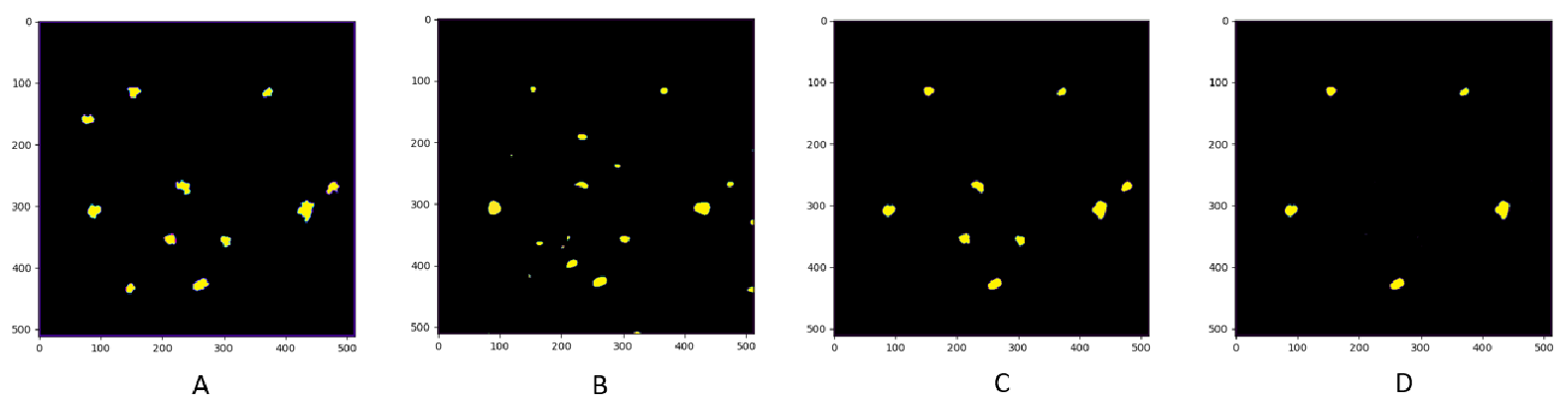}
\caption{A: clusters that have been selected by DBSCAN; B: clusters selected by 2 sigma analysis; C: Plane logical AND on a 2D slice; D: Logical AND filtered with tha superimposition over the third dimension.}\label{fig:clusters}
\end{figure}

The fourth step is to check if the identified clusters are present in the adjacent slices. If this condition is not satisfied, the cluster is attributable to noise perturbation and discarded. Fig. \ref{fig:clusters} schematizes the output matrices of the procedure described above, step-by-step.\\ 

Once the remaining clusters are labeled as belonging to a specific insert, a linear regression on their centers of massallows to identify the propagation axis of the tridimensional inserts. \\

\begin{figure}[h!]
\centering
\includegraphics[scale=0.5]{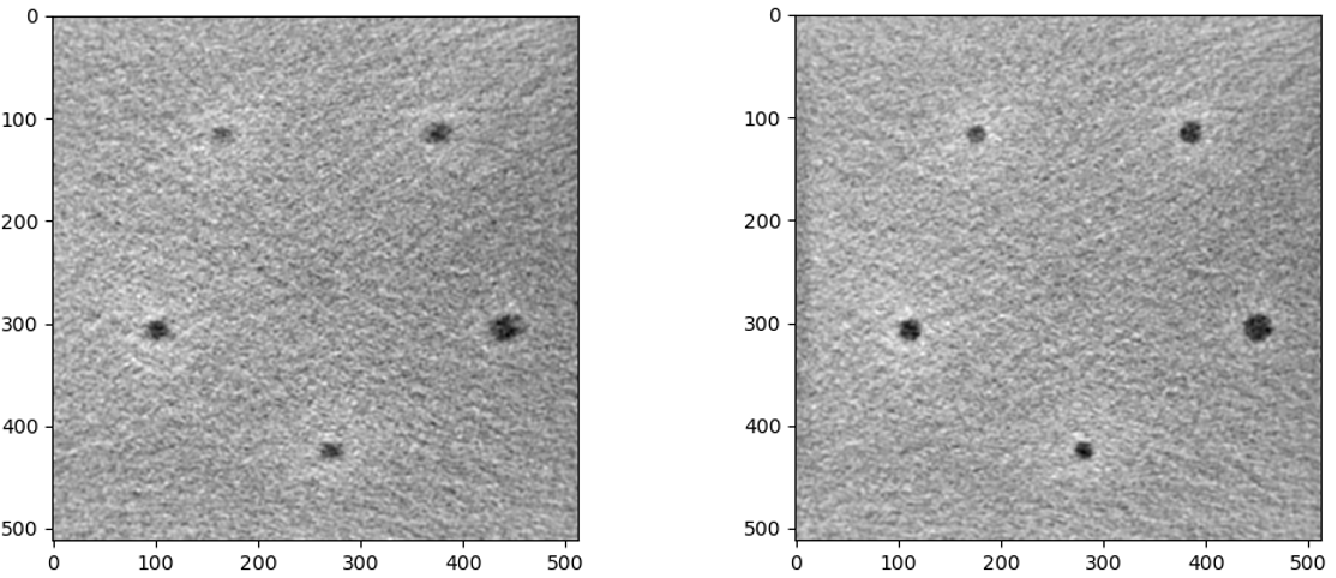}
\caption{A comparison of a super resolution image obtained with the simple sum over Z-axis (on the left) and the one obtained after proper translation of the CT slices (on the right).}\label{fig:super_res}
\end{figure}

The knowledge of the propagation axis allows the proper displacement of the slices so that the center of mass of a cluster within a slice is aligned in all the other slices. At this point, a super-resolution image can be generated, that is an image obtained by the summation of all the aligned slices. Our study made the assumption that the axis was fixed in space; however, in a clinical context, this assumption would not hold, and displacements should be considered independently for each slice. The correct slices displacement allows to obtain, in the final summed image, a substantial increase of the contrast-noise ratio (CNR). The final result is a much sharper image (actually, a super resolution image) allowing a much easier detection of very low contrast signals. The super-resolution images obtained after proper slices alignment and summation were compared with those generated by summation along the Z axis of the original slices. An example can be seen in figure \ref{fig:super_res}. \\

\begin{figure}[h!]
\centering
\includegraphics[scale=0.5]{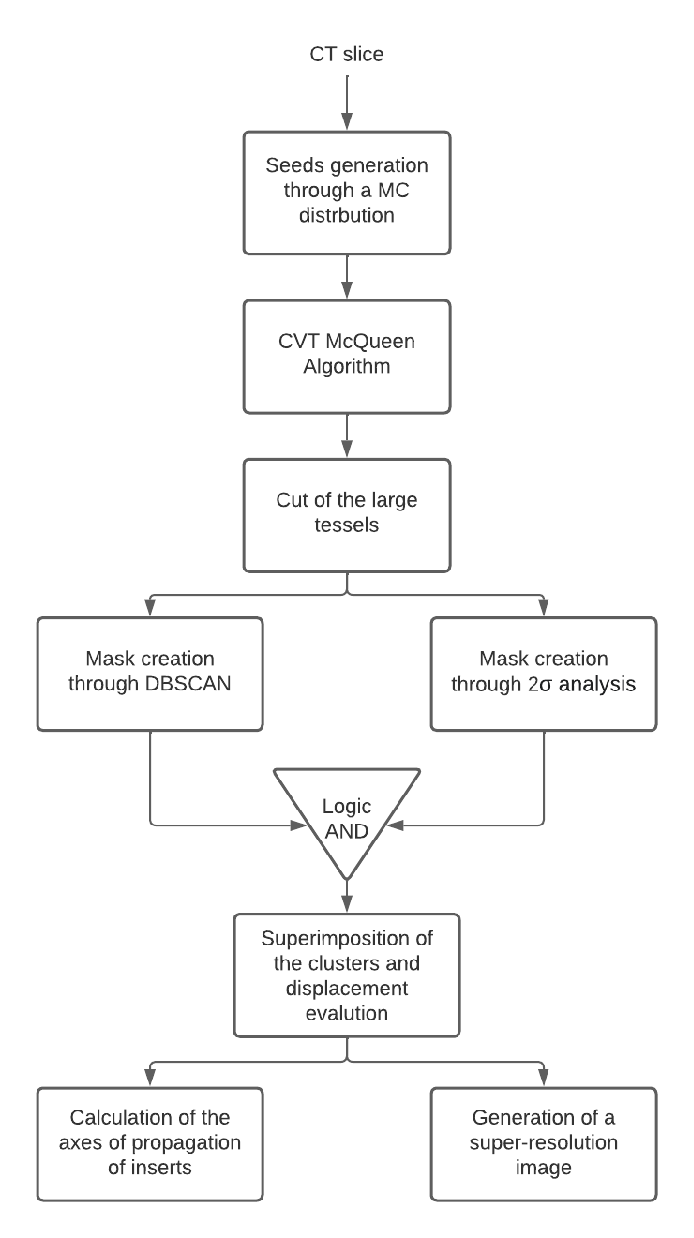}
\caption{Block diagram of the algorithm described in the section The Algorithm}\label{fig:blocchi}
\end{figure}

A block diagram of the whole process is shown in the figure \ref{fig:blocchi}.

The results of the algorithm were obtained in computational time around 5-8 minutes in an ordinary laptop – MSI Prestige 14 Evo with 3.00GHz Intel Core i7-1185g7. The full code was developed on python 3.9.6.

\section{Results}

The algorithm based on Centroidal Voronoi Tessellation is able to correctly identify the clusters belonging to low-contrast regions, obtaining performances that do not suffer from certain deficiencies that may occur to a human observer, such as: tiredness, distraction or non-optimal reporting conditions.\\
The detection rate of the inserts in the single CT slices is measured as the percentage of the number of inserts detected in all the acquired slices. The results show that the algorithm was able to classify the clusters as being part of the various insert at least in 60\% of the cases: this low detection rate is relative to the insert with only 8 HU of contrast difference compared to PMMA at the lowest $CTDI_{vol}$, and 0.6 mm of slice thickness, outermost in clinical practice. On the contrary for the contrast of 25HU and 35HU the detection rate is very good, up to a detection rate of 91\% for the slice thickness of 1 mm and a higher $CTDI_{vol}$ (table \ref{tab:res1}).\\

\begin{table}[]
\centering
\begin{tabular}{ccc}
\hline
Slice Thickness & CTDI (mGy) & Average Detection Rate \\ \hline
0.6 mm          & 3.54       & 60\% \\ \hline
0.6 mm          & 7.05       & 82\% \\ \hline
1 mm            & 3.54       & 81\% \\ \hline
1 mm            & 7.05       & 93\% \\ \hline
\end{tabular}
\caption{Results of the detection rate: the percentages refers to the average of all five inserts in the 2D slices.}
\label{tab:res1}
\end{table}

A qualitative analysis of super-resolution images shows how the sum of the slices along the Z axis produces pronounced ovoid-shaped inserts when compared to the more circular ones obtained by summing along the actual axis identified by our algorithm which therefore allows to reproduce a shape more similar to the real one. The second emergent effect in the images obtained by the summation along the propagation axis, is the sharpening of the signal; de facto, the peripheral regions do not exhibit an attenuated signal intensity, as opposed to the sum along the z-axis, where the inclusion of noisy areas can impact the signal strength. This behavior, was quantified through the analysis of the Contrast to Noise Ratio (CNR), calculated on a crown-shaped ROI surrounding the insert with an area equal to insert itself. CNR shows an increase from a minimum of 20\% up to a maximum of 32\%, as reported in the table \ref{tab:res2}.\\

\begin{table}[]
\centering
\begin{tabular}{cccc}
\hline
Slice Thickness & CTDI (mGy) & CNR   \\ \hline
0.6 mm          & 3.54       & +20\% \\ \hline
0.6 mm          & 7.05       & +25\% \\ \hline
1 mm            & 3.54       & +30\% \\ \hline
1 mm            & 7.05       & +32\% \\ \hline
\end{tabular}
\caption{Increase of the CNR in the super-resolution image obtained by summing along the actual axis of the insert instead of summing along the z axis.}
\label{tab:res2}
\end{table}

\begin{figure}[h!]
\centering
\includegraphics[scale=0.5]{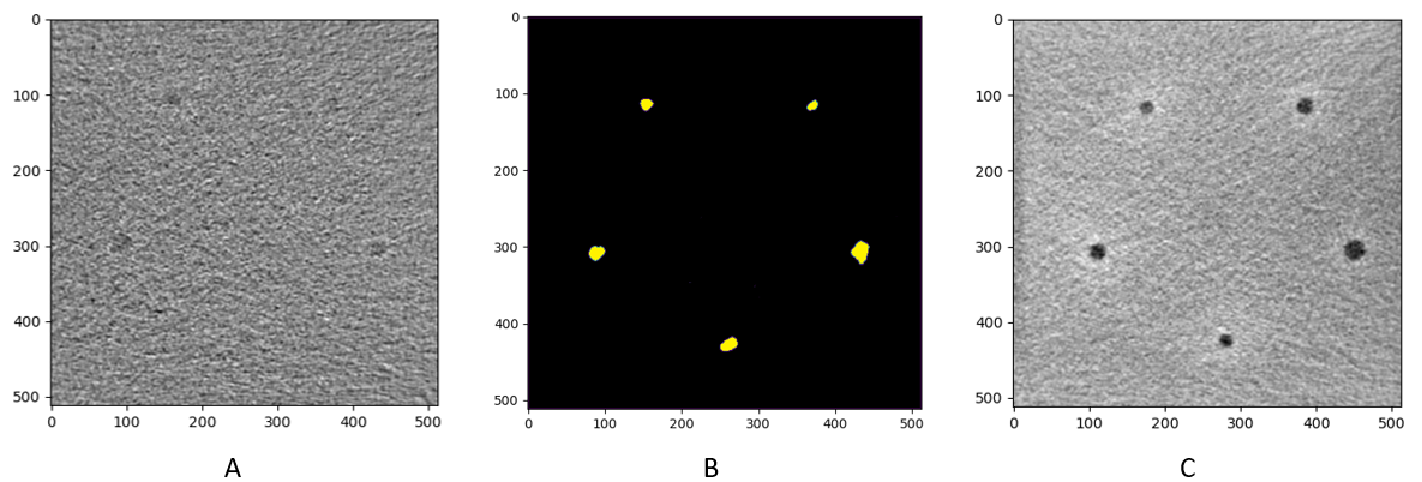}
\caption{A: Original CT 2D slice; B: cluster selected for that slice; C: Super-resolution image obtained for that acquisition.}\label{fig:results}
\end{figure}

These results in super-resolution images can be obtained only if the propagation axis is correctly identified, in fact a summation along an incorrect axis would lead to sub-optimal results, if not even worse than the reconstruction along the z-axis. 
The displacement, of each center of mass respect to the previous one, was correctly identified with an error of a few percent as reported in table \ref{tab:res3}. The results from both the single-slice cluster identification and the super-resolution are displayed in the figure \ref{fig:results}.

\begin{table}[]
\centering
\begin{tabular}{cccc}
\hline
Slice Thickness & CTDI (mGy) & Real Displacement & Calculated Displacement   \\ \hline
0.6 mm & 3.54 & X = 0.172 mm;  & X = 0.172 ± 0.006 mm;\\ 
       &      & Y = 0.000 mm;  & Y = 0.00 ± 0.01 mm\\ \hline
0.6 mm & 7.05 & X = 0.172 mm;  & X = 0.171 ± 0.008 mm; \\
       &      & Y = 0.000 mm;  & Y = 0.004 ± 0.0041 mm\\ \hline
1 mm   & 3.54 & X = 0.287 mm;  & X = 0.301 ± 0.008 mm; \\
       &      & Y = 0.000 mm;  & Y = 0.01 ± 0.01 mm\\ \hline
1 mm   & 7.05 & X = 0.287 mm;  & X = 0.289 ± 0.01 mm; \\
       &      & Y = 0.000 mm;  & Y = 0.01 ± 0.03 mm\\ \hline
\end{tabular}
\caption{Results of the shift calculated by the algorithm respect to the real shift}
\label{tab:res3}
\end{table}

\section{Conclusions}

In this work a newly developed algorithm, based on the Centroidal Voronoi Tessellation, was developed and tested to perform detection of low contrast objects in CT images obtained by acquiring a specifically designed PMMA phantom under different $CTDI_{vol}$ settings and reconstrion slice thicknesses. The results showed a high detection rate, meaning that it is possible identify lesions of clinical interest even at very low contrast.
The study also demonstrates the effectiveness of a novel super resolution image production obtained by means of a targeted summation of the CT slices to produce much sharper definition of the low contrast inserts. \\
The algorithm showed its limitations with the insert used as a stress test that is (contrast of 8HU, 5mm of diameter, 0.6mm of slice thickness and 3.54mGy of $CTDI_{vol}$) it failed to identify this signal source at all. When considering slice thickness and $CTDI_{vol}$ values that are most representative of clinical practice, the same insert demonstrated the potential for identifying very low-contrast lesions using the proposed technique. \\
In summary, the results of this study demonstrated the potentiality of our procedure for further developments and applications to detect clinical lesions in homogenous anatomic districts, such as the liver. We are encouraged by the results obtained that we plan to ask the authorization of the ethics committee for the application of our procedure to clinical tasks with the aim of investigating the possibility of an early diagnosis of some liver diseases such as the hepatocellular carcinoma.We selected this particular pathology because it represents the described scenario of low-contrast lesions within a homogenous anatomical region.

\acknowledgments
The author acknowledge Dr. Elena Cantoni and Dr. Angela Muggiolu for the long discussions and support in optimizing the algorithm, further thanks go to the whole team of the Department of Physics and Astronomy of the University of Florence and of the Medical Physics Department of the Careggi University Hospital who suppported this work.

\end{document}